\documentclass{article}

\usepackage[usenames,dvipsnames,svgnames]{xcolor}

\usepackage{hyperref}
\hypersetup{
    colorlinks=true,
    citecolor=MidnightBlue,
    urlcolor=Bittersweet,
}

\usepackage[margin=1in]{geometry}

\usepackage{float}
\usepackage{bbm}
\usepackage{bm}
\usepackage[normalem]{ulem}
\usepackage{extarrows}
\usepackage{enumitem}
\usepackage{comment}
\usepackage{amsthm,amsmath,amsfonts,amssymb}
\usepackage{booktabs}
\usepackage{algorithm,algpseudocode} 

\usepackage{cleveref}

\usepackage{graphicx}
\graphicspath{figure/}
\usepackage{subfigure}
\usepackage{caption}

\usepackage{algorithm}
\usepackage{algorithmicx}
\usepackage{algpseudocode}
\usepackage{url}
\usepackage{verbatim}
\usepackage{makecell}
\usepackage{tikz}
\usepackage{cite}
\usetikzlibrary{spy}
\usepackage{multirow}
\usepackage[super,sort,comma]{natbib}

\newcommand{\x}{\mathbf{x}}	
\newcommand{\y}{\mathbf{y}}
\newcommand{\A}{\mathbf{A}}
\newcommand{\B}{\mathbf{B}}

\newcommand{\z}{\mathbf{z}}
\newcommand{\W}{\mathbf{W}}
\renewcommand{\P}{\mathbf{P}}
\newcommand{\omg}{\mathbf{\Omega}}

\newcommand{\Z}{\mathbf{Z}}
\newcommand{\0}{\mathbf{0}}

\newcommand{\I}{\mathbf{I}}

\newcommand{\s}{\mathbf{s}}
\newcommand{\Rsf}{\mathsf{S}}
\newcommand{\g}{\mathbf{g}}
\newcommand{\h}{\mathbf{h}}
\newcommand{\ze}{\bm{\zeta}}

\newcommand{\R}{\mathbf{R}}
\newcommand{\diag}{\mathsf{diag}}
\newcommand{\U}{\mathbf{U}}	
\newcommand{\V}{\mathbf{V}}
\newcommand{\Sig}{\mathbf{\Sigma}}

\renewcommand{\r}{\mathbf{r}}
\renewcommand{\b}{\mathbf{b}}
\renewcommand{\H}{\mathbf{H}}
\newcommand*\samethanks[1][\value{footnote}]{\footnotemark[#1]}

\begin{document}
\title{Multi-layer Clustering-based Residual Sparsifying Transform for Low-dose CT Image Reconstruction}

\author{
Xikai Yang\thanks{University of Michigan - Shanghai Jiao Tong University Joint Institute, Shanghai Jiao Tong University, Shanghai 200240, China, yong.long@sjtu.edu.cn} , 
Zhishen Huang\thanks{Department of Computational Mathematics, Science and Engineering (CMSE), Michigan State University, East Lansing, MI 48824, USA} ,
Yong Long\samethanks[1] , 
Saiprasad Ravishankar\samethanks[2] \thanks{Department of Biomedical Engineering, Michigan State University, East Lansing, MI 48824, USA}
}

\maketitle

\begin{abstract}
	\noindent {\bf Purpose:} The recently proposed sparsifying transform models incur low computational cost and have been applied to medical imaging. Meanwhile, deep models with nested network structure reveal great potential for learning features in different layers. In this study, we propose a network-structured sparsifying transform learning approach for X-ray computed tomography (CT), which we refer to as multi-layer clustering-based residual sparsifying transform (MCST) learning. 
	The proposed MCST scheme learns multiple different unitary transforms in each layer by dividing each layer's input into several classes. 
	We apply the MCST model to low-dose CT (LDCT) reconstruction by deploying the learned MCST model into the regularizer in penalized weighted least squares (PWLS) reconstruction.\\
	
	\noindent{\bf Methods:}
	The proposed MCST model combines a multi-layer sparse representation structure with multiple clusters for the features in each layer that are modeled by a rich collection of transforms.
	We train the MCST model in an unsupervised manner via a block coordinate descent algorithm. Since our method is patch-based, the training can be performed with a limited set of images.
For CT image reconstruction, we devise a novel algorithm called PWLS-MCST by integrating the pre-learned MCST signal model with PWLS optimization.\\

\noindent{\bf Results:} We conducted LDCT reconstruction experiments on XCAT phantom data and Mayo Clinic data. We trained the MCST model with 2 (or 3) layers and with 5 clusters in each layer. The learned transforms in the same layer showed rich features while additional information is extracted from representation residuals. Our simulation results demonstrate that PWLS-MCST achieves better image reconstruction quality than the conventional FBP method and PWLS with edge-preserving (EP) regularizer. It also outperformed recent advanced methods like PWLS with a learned multi-layer residual sparsifying transform prior (MARS) and PWLS with a union of learned transforms (ULTRA), especially for displaying clear edges and preserving subtle details.\\

\noindent{\bf Conclusions:} In this work, a multi-layer sparse signal model with a nested network structure is proposed. We dub this novel model as the MCST model that exploits multi-layer residual maps to sparsify the underlying image and clusters the inputs in each layer for accurate sparsification.
We presented a new PWLS framework with a learned MCST regularizer for LDCT reconstruction. Experimental results show that the proposed PWLS-MCST provides clearer reconstructions than several baseline methods.  
\end{abstract}


\section{Introduction}


Natural signals such as images, audio and video often demonstrate sparsity in some representation domain such as wavelets~\cite{pati:93:omp}, Fourier transform, etc. Dictionary learning methods can extract features from natural signal data and represent the natural signal in a sparse format.
Such sparse representation methods have been used in many areas like image denoising~\cite{elad:06:idv}, super-resolution~\cite{yang:10:isr}, inpainting~\cite{julien:08:src}, medical image reconstruction~\cite{zhang:17:tbd},  etc.

There are two typical approaches to find a sparse representation of a signal: the synthesis model and the analysis model. Let $\y$ be the measurement outcome of signal $\x$ with measurement operator $\Phi$. For synthesis models, a signal $\x \in \mathbb{R}^n$ is encoded as a combination of column vectors in a synthesis dictionary $\omg_S \in \mathbb{R}^{n\times m}$ by assuming $\x = \omg_S\z$, where $\z \in \mathbb{R}^{m}$ contains only a few nonzero elements, and then the data consistency is enforced simultaneously in the form of $\min_\x \|\y-\Phi\x\|_2$. For analysis models, a signal is assumed to be sparse in an (overcomplete) transform domain with $\omg_A\x=\z$, where $\omg_A \in \mathbb{R}^{m\times n}$ is referred to as the analysis dictionary, and meanwhile one in parallel enforces the data consistency $\min_\x \|\y-\Phi\x\|_2$.

The sparse coding step is NP-hard, and the algorithms to learn synthesis or analysis dictionaries are computationally expensive such as K-SVD~\cite{aharon:06:ksa}, Analysis K-SVD~\cite{rubinstein:13:aka}, NAAOL~\cite{yaghoobi:13:coa}, etc. 
The sparsifying transform model manifests advantage in terms of low computational cost for computing the sparse coding of signals~\cite{ravishankar:15:lst}. By modeling $\omg\x=\z+\eta$ where $\eta$ indicates an error term, $\x$ is approximately sparse in the transform domain. Such a sparsifying transform model~\cite{ravishankar:13:lst} can be regarded as a generalized analysis model. 
For instance, Ravishankar et al.\ proposed to learn doubly sparse transforms for signals~\cite{ravishankar:13:lds}, where the sparsifying transform is a product of two different transforms, a sparse transform and a fast, analytical transform. Other works also developed efficient blind compressed sensing algorithms~\cite{sabsam} for joint sparsifying transform learning and image reconstruction for MRI.

Recent years have witnessed the growing development of deep learning methods. A typical method is to employ deep neural networks (DNN) and train an end-to-end model between the input and the output. For example, Jin et al. proposed the FBPConvNet based on the U-net framework~\cite{jin:17:dcn}, which constructs a mapping from noisy low-dose CT reconstruction images to high-quality images. Chen et al. combined the autoencoder and deconvolution network and proposed the RED-CNN framework for low-dose CT imaging~\cite{chen:2017:redCNN}. Despite that DNN-based methods show potential for capturing underlying features of big data, a large amount of (often paired) training data is indispensable to train a feasible network. In addition, an explicit understanding of the modeling or representation captured in the network is currently lacking.


In the past few years, researchers imitated the nested network structure and created more complex sparse signal models. For example, Tang et al. proposed a Deep Micro-Dictionary Learning and Coding Network, where classic convolutional layers are replaced with dictionary learning and feature encoding operations~\cite{tang:19:dmd}.
Singhal et al.~\cite{Singhal:18:mmt} presented a deep dictionary learning network (DDL) for image classification by minimizing the reconstruction error after going through multiple DL feature encoding steps. Ravishankar et al. exploited multi-layer extension for sparsifying transform model, which enables sparsifying an input image layer by layer~\cite{ravishankar:18:lml}.
Other works proposed to improve signal models by combining multiple signal modules in a parallel structure~\cite{wen:14:sos,zheng:18:pua}. 
Wen et al. extended the single sparsifying transform model to a union of sparsifying transforms model, where different inputs are assigned to different clusters and transforms are learned with respect to each cluster~\cite{wen:14:sos}. 

Several applications of sparsifying transform (ST) learning 
and various extensions have been demonstrated in the field of medical image reconstruction such as magnetic resonance image reconstruction and computed tomography (CT) image reconstruction~\cite{zheng:18:pua,yang:20:mars}. The conventional method for regular-dose CT image reconstruction is the analytical filtered back-projection (FBP)~\cite{feldkamp:84:pcb}. However, with reduced/low X-ray dose, severe artifacts and noise can degrade the quality of the reconstructed image. 
A class of reconstruction methods takes the physics model of the imaging system into account along with statistical models of measurements and noise and often simple object priors. These methods are referred to as model-based image reconstruction methods (MBIR)~\cite{MBIR_review_21}.

The penalized weighted least squares (PWLS) approach is a common MBIR method, whose loss function includes a weighted quadratic data-fidelity term and a regularizer term. The regularizer is conventionally based on hand-crafted models and priors~\cite{sauer:93:alu,thibault:06:arf,thibault:07:atd,cho:15:rdf}.
Previous studies manifested a promising result by introducing learned sparsifying transform learning into the PWLS scheme~\cite{pfister:14:mbi,pfister:14:trw,pfister:14:ast,chun:17:esv}.  The sparsifying transform learned in an unsupervised manner is incorporated into a prior term, which is equivalent to involving effective information extracted from the training slices into the process of image reconstruction. Moreover, using multiple sparsifying transforms increases the number of obtained features~\cite{zheng:18:pua,yang:20:mars} and further boosts the performance of reconstruction algorithms.

In this study, we propose a network-structured sparsifying transform learning module, 
which we refer to as the \textit{Multi-layer Clustering-based residual Sparsifying Transform} (MCST) model. We extended the original sparsifying transform model in two aspects. On the one hand, the multi-layer structure of MCST model enables gaining potential features by exploiting residual maps layer by layer.
On the other hand, the clustering operation in each layer implements an attention-like mechanism 
by grouping input patches. We propose a pipeline for training the MCST model and show how to update the variables in this pipeline. In addition, we have applied the proposed learned MCST model to LDCT reconstruction through a PWLS optimization scheme. We named the new LDCT reconstruction algorithm \textit{PWLS-MCST}. Experimental results demonstrate that PWLS-MCST not only provides better image reconstruction quality than conventional methods like FBP and PWLS method with the non-adaptive edge-preserving (EP) regularizer, but also outperforms several advanced methods such as PWLS-ULTRA~\cite{zheng:18:pua} and PWLS-MARS~\cite{yang:20:mars} in terms of RMSE and SSIM.

The rest of this paper is organized as follows. In Section~\ref{chap:MCST_method}, we introduce the MCST model and describe its application to image reconstruction in the LDCT setting. The experimental results are presented in Section~\ref{chap:MCST_experiment}. We summarize our conclusions in Section~\ref{chap:MCST_conclude}.

\section{Methods}\label{chap:MCST_method}

\subsection{Model description}
\begin{figure}[h]
	\centering
	\includegraphics[width=1.0\textwidth]{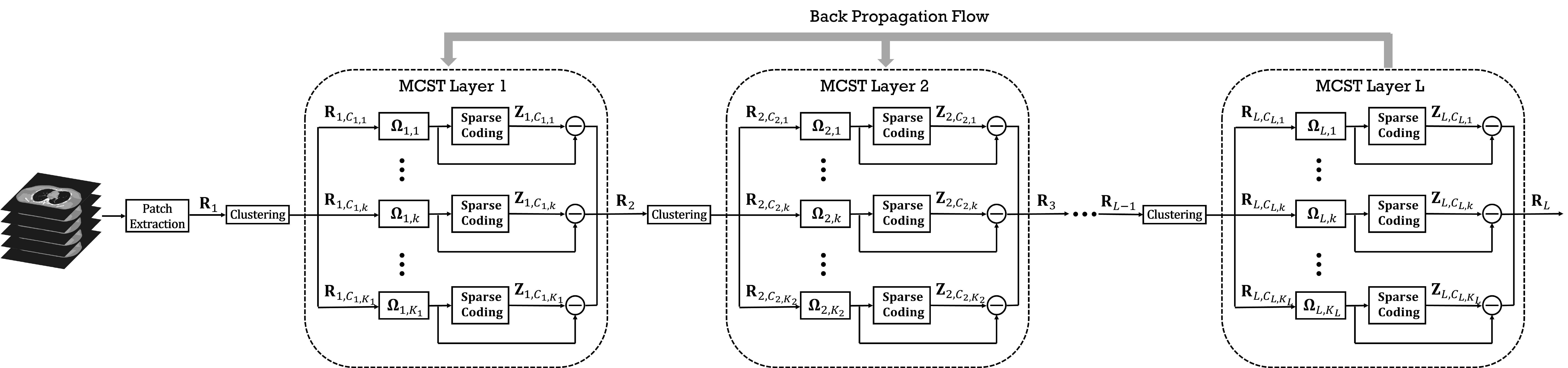}
	\caption{Illustration of multi-layer clustering-based sparsifying transform model. $\omg_{l,k}$ denotes the transform  corresponding to class $k$ in layer $l$. $\R_{l,{C_{l,k}}}$, $\Z_{l,{C_{l,k}}}$ represent residual map and sparsecode submatrices in layer $l$ with column indices $i \in C_{l,k}$ respectively, where $C_{l,k}$ is a set containing indices of clustered variables from class $k$ in the $l$-th layer.}
	\label{fig:MCST_structure}
\end{figure}

\begin{figure}[h!]
	\centering
	\includegraphics[width=0.95\textwidth]{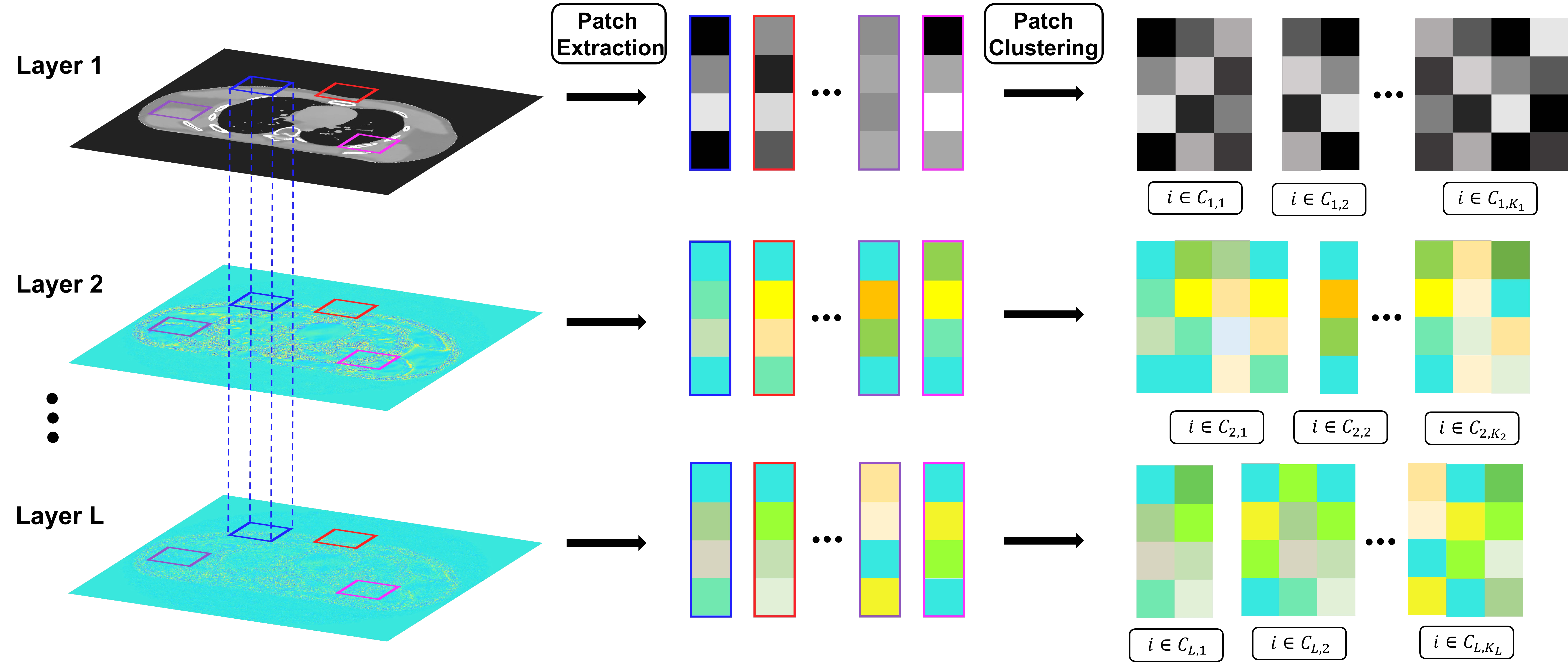}
	\caption{Illustration of the patch extraction and patch clustering processes on an image across multiple layers. For patches with the same index, even though they are in different layers, the location of the included pixels in the image is the same. Each patch is vectorized for further clustering processing after extraction.}
	\label{fig:patch_extraction}
\end{figure}

In this section, we give a mathematical description of the MCST model and introduce its framework. Figure~\ref{fig:MCST_structure} shows the structure of the proposed multi-layer clustering-based residual sparsifying transform (MCST) model. Table~\ref{table:Notation_sys_MCST} in Appendix gives a detailed description of the notation used in our methodology.
Matrix $\omg_{l,k}$ denotes the transform in class $k$ in layer $l$. $\R_{l,{C_{l,k}}}$ and $\Z_{l,{C_{l,k}}}$ represent residual map and sparse code submatrices respectively, in layer $l$ with column indices $i \in C_{l,k}$, where $C_{l,k}$ is a set containing indices of clustered variables of class $k$ in the $l$-th layer.
In particular, each column in $\R_1$ is a vectorized overlapping image patch extracted from the training dataset (of images) with an appropriate patch stride. 
Figure~\ref{fig:patch_extraction} gives an illustration of the patch extraction and patch clustering processes on an image across multiple layers. For patches with the same index, even though they are in different layers, the location of the included pixels in the image is the same. Each patch is vectorized for further clustering processing after extraction.
The proposed MCST model extends the original sparsifying transform model in two dimensions to strengthen the model's representation ability. The first type of extension is increasing the depth of the model. We borrow the design idea of the recent MARS~\cite{yang:20:mars} model. 
Multi-layer structure enforces the filtered residuals (the difference between the transformed input signals and their sparse representation) to be further sparsified layer by layer. 
Each layer keeps partial information or captures effective features from the input. 
The second type of improvement is changing the width of the MCST model. Similar to the popular attention mechanism, for different parts of the input signal in each layer, we learn sparsifying transform modules separately. Since our method is patch-based, a clustering operation is necessary to assign each patch into a specific class. Moreover, the relative order of the patches cannot be disrupted by the clustering operation, which requires us to remember the index of each patch at each layer and combine them in the same order before the next clustering.

The objective function corresponding to the MCST model's training is as follows. Due to the complexity of the proposed method, we present the notation system used in \eqref{eq:P2} in Table~\ref{table:Notation_sys_MCST}.
\begin{align}\label{eq:P2}
	& \min_{\{\omg_l,\Z_l, C_{l,k}\}}   \sum_{l=1}^L \sum_{k=1}^{K_l} \sum_{i\in C_{l,k}} \bigg\{  \|\omg_{l,k}\mathbf{r}_{l,i} - \z_{l,i}\|_2^2 + \eta_l^2\|\z_{l,i}\|_0  \bigg\} \nonumber\\
	&\mathrm{s.t.} \quad \begin{cases}
		\mathbf{r}_{l+1,i} = \omg_{l,k}\mathbf{r}_{l,i} - \z_{l,i}, &  1\leq l\leq L-1,\\
		\omg_{l,k} \omg_{l,k}^\top = \omg_{l,k}^\top \omg_{l,k}  = \I,  &\forall l,k.
	\end{cases}
	\tag{P0}
\end{align}

\subsection{Algorithm for model training and image reconstruction}
\begin{figure}[h!]
	\centering
	\includegraphics[width=1.0\textwidth]{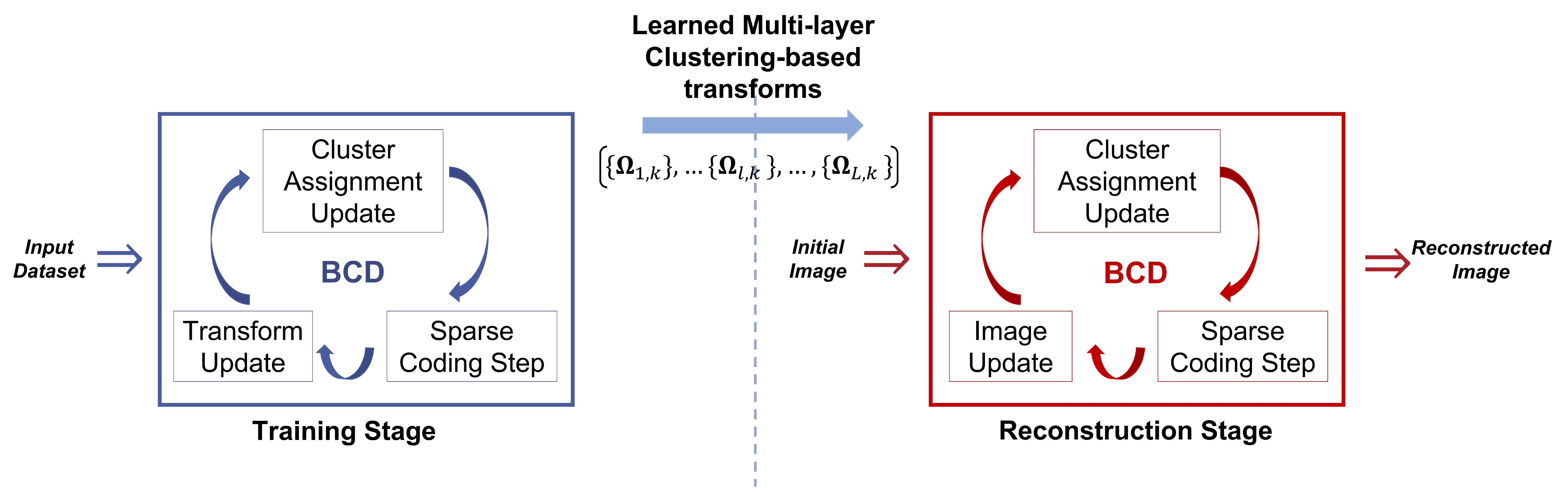}
	\caption{Overview of the proposed PWLS-MCST algorithm for low-dose CT reconstruction.}
	\label{fig:MCST_overview}
\end{figure}

Figure~\ref{fig:MCST_overview} gives an overview of the algorithms for model training and low-dose CT reconstruction. The whole process is divided into two stages. In the process of model training, \eqref{eq:P2} is solved by employing a block coordinate descent (BCD) method. We propose an iterative algorithm involving cluster assignment update, transform update, and sparse coding steps. Note that we use unlabelled images to train the proposed MCST signal model. A similar cyclical updating strategy is applied to the image reconstruction stage as well. Instead of updating the transform in each iteration, for the reconstruction stage, an image update step is included into the BCD algorithm. We designed the prior function (regularizer) incorporating the learned MCST model to harness the effective information obtained in the training stage. 

\subsubsection{Algorithm for MCST model training}\label{sec:alg_mcst_train}
The formulation for training the MCST model is given as in \eqref{eq:P2}. Since \eqref{eq:P2} is nonconvex, similar to the recent MARS work~\cite{yang:20:mars},
we apply the BCD algorithm to solve this problem, which takes an iterative updating strategy among different variables to be optimized. 
Algorithm~\ref{alg: mcst_train} shows the full MCST learning pipeline. We update variables in MCST by looping over layers. The specific solving procedure is presented in the following steps:
\begin{itemize}
	\item [a)] Cluster assignment update in $l$-th layer 
	
	In this step, we find the cluster assignment for every patch in the $l$th layer by solving the following optimization problem \eqref{eq:update_Cl}. Random initialization is employed for cluster assignment in the first iteration. When updating the cluster assignment in the $l$-th layer, the patch assignments in the deeper layers $\{k(i,s), s>l\}$ are  fixed.
	\begin{equation}\label{eq:update_Cl}
		\begin{aligned}
			&\widehat{k}(i,l) = \arg\min_{1\leq k \leq K_l}\|\omg_{l,k(i,l)}\r_{l,i} - \z_{l,i}\|_2^2 + \sum_{s=l+1}^L \|\omg_{s,k(i,s)}\r_{s,i} - \z_{s,i}\|_2^2 + \eta_l^2\|\z_{l,i}\|_0
			\\
			& \quad \mathrm{s.t.} \quad 
				\r_{s,i} = \omg_{s-1,k(i,s-1)}\r_{s-1,i} - \z_{s-1,i}, \quad l+1\leq s\leq L-1.
		\end{aligned}
	\end{equation}
	
	\paragraph{Interpretation} The objective function of optimization problem \eqref{eq:update_Cl} consists of two parts: sparse encoding residual and an $\ell_0$ penalty to enforce sparsity on the encoded representation. When computing the cluster assignment of a patch $i$ in the $l$-th layer, which is denoted as $k(i,l)$, the transforms in later layers ($\omg_{s,\cdot}$ where $s>l$) still play a role. This is because the residuals $\r_{s,i}$ of the patch $i$ at later layers ($s>l$) depend on transforms ($\omg_{l,\cdot}$) and the residual ($\r_{l,i}$) in the $l$-th layer, which is reflected in the constraint. 
	
	To solve problem \eqref{eq:update_Cl}, we start at layer $l$ and proceed to later layers. To determine $\widehat{k}(i,l)$, one only needs to enumerate the values of the objective function with $k\in[K_l]$, and choose the index $k$ that gives the minimal objective function value. We repeat the solution process across all layers until the cluster assignment no longer changes. 
	
	
	\item [b)] Sparse coding step for $\Z_l$
	
	The objective to update $\Z_l$ in each iteration is shown in \eqref{eq:mcst_sub_pro_Z}. With the cluster assignment for every patch $\{C_{l,k},1\leq l\leq L, 1\leq k \leq K_l\}$ and all transforms $\{\omg_{l,k}, 1\leq l\leq L, 1\leq k \leq K_l\}$ and sparse coefficient maps except the $l$th layer $\{\Z_k, \forall k \neq l\}$ fixed, the objective function in \eqref{eq:P2} becomes
	
	\begin{equation}\label{eq:mcst_sub_pro_Z}
		\begin{aligned}
			&\min_{\Z_l} \sum_{s=l}^L \sum_{k=1}^{K_s} \sum_{i\in C_{s,k}} \bigg\{  \|\omg_{s,k}\r_{s,i} - \z_{s,i}\|_2^2\bigg\} + \eta_l^2\|\Z_{l}\|_0.
		\end{aligned}
	\end{equation}	
	
	We decompose \eqref{eq:mcst_sub_pro_Z} into \eqref{eq:sub_pro_Z_i} which concerns only one patch, and solve 
	for all
	$\{\z_{l,i},1\leq i \leq N\}$ in parallel. 
	\begin{equation}\label{eq:sub_pro_Z_i}
		\begin{aligned}
			&\min_{\z_{l,i}} \sum_{s=l}^L \bigg\{  \|\omg_{s,k(i,s)}\r_{s,i} - \z_{s,i}\|_2^2\bigg\} + \eta_l^2\|\z_{l,i}\|_0   
		\end{aligned}
	\end{equation}	
	
	To solve~\eqref{eq:sub_pro_Z_i},	
	we fix the base layer index as $l$. Recall that the residuals $\r_s$ in later layers $(s>l)$ depends on the residual $\r_l$ and the encoding vector $\z_l$ in layer $l$. Therefore, we need to peel off this dependence to find the solution for the encoding vector in layer $l$. We use the unitary property of the transforms to disentangle the dependence. Consider the encoding discrepancy term in a layer $s$ where $s=l+1,l+2,\cdots, L$:
	\begin{align}
		\label{eq:disentangle}
		\|\omg_{s,k(i,s)} \r_{s,i_{l}} - \z_{s,i}\|_2 
		&= \|\underbrace{\omg_{s,k(i,s)}^\top \omg_{s,k(i,s)}}_{\mathbf{I}} \r_{s,i} - \omg_{s,k(i,s)}^\top\z_{s,i}\|_2 \nonumber \\
		&= \|\r_{s,i} - \omg_{s,k(i,s)}^\top\z_{s,i}\|_2 \nonumber\\
		&= \|\underbrace{\omg_{s-1,k(i,s-1)} \r_{s-1,i} - \z_{s-1,i}}_{\r_{s,i}} - \b_{i}^{(s-1)\leftarrow s} \|_2 \nonumber\\
		&= \| \r_{s-1,i} - \underbrace{\omg_{s-1,k(i,s-1)}^\top\z_{s-1,i} - \omg_{s-1,k(i,s-1)}^\top \b_{i}^{(s-1)\leftarrow s}}_{\b^{(s-2)\leftarrow s}_{i}} \|_2 \nonumber\\
		&= \| \omg_{s-2,k(i,s-1)} \r_{s-2,i} - \z_{s-2,i} - \b_{i}^{(s-2)\leftarrow s} \|_2 \nonumber\\
		&= \cdots \cdots \nonumber\\
		&= \| \omg_{l,k(i,l)} \r_{l, i} - \z_{l,i} - \b_{i}^{l\leftarrow s} \|_2,
	\end{align}
	where we define the back-propagation information vector from the $q$-th to the $p$-th layer for the patch $i$ as
	\begin{equation}\label{eq:B_pq}
		\begin{aligned}
			\b^{p\leftarrow q}_{i} &= \omg_{p+1,k(i,p+1)}^\top \z_{p+1,i} + \omg_{p+1,k(i,p+1)}^\top\omg_{p+2,k(i,p+1)}^\top\z_{p+2,i} + \cdots + \\
			&\quad\quad \omg_{p+1,k(i,p+1)}^\top\omg_{p+2,k(i,p+2)}^\top \cdots \omg_{q,k(i,q)}^\top \z_{q,i}
			\\
			&=\sum_{s=p+1}^q \bigg( \prod_{m=p+1}^s \omg_{m,k(i,m)}^\top \bigg)\z_{s,i}.
		\end{aligned}
	\end{equation}
	Consequently, the objective function in the optimization problem \eqref{eq:sub_pro_Z_i} can be re-written as \[ 
	\min_{\z_{l,i}} \|\omg_{l,k(i,l)} \r_{l, i} - \z_{l,i} \|^2_2 + \sum_{s=l+1}^L \|\omg_{l,k(i,s)} \r_{l, i} - \z_{l,i} - \b_{i}^{l\leftarrow s}\|^2_2 + \eta_l^2\|\z_{l,i}\|_0.
	\]
	
	The solution of optimal $\z_{l,i}$ to the optimization problem above is thus 
	\begin{equation}\label{eq:sol_Z_l}
		\widehat{\z}_{l,i}=
		\begin{cases}
			H_{\eta_l/\sqrt{L-l+1}} \bigg( \omg_{l,k(i,l)}\r_{l,i} - \frac{1}{L-l+1}\sum_{s=l+1}^L \b^{l\leftarrow s}_{i} \bigg), &  1\leq l \leq L-1,\\
			H_{\eta_L} (\omg_{L,k(i,L)} \r_{L,i}), & l=L,
		\end{cases}
	\end{equation}
	The solution form depends on the specific layer $l$, where optimization is done in \eqref{eq:mcst_sub_pro_Z}.

	\item [c)] Transform update for $\{\omg_{l,k},1\leq k \leq K_l\}$
	
	
	
	To update the transforms in the $l$-th layer, we leave out quadratic terms with layer depth less than $l$ and all $\|\cdot\|_0$ terms which are unrelated to the target optimization variables. By assembling the column vectors $\mathbf{r}_{s,i}$,  $\z_{s,i}$ into $\mathbf{R}_{s, C_{s,k}}$ and $\Z_{s , C_{s,k}}$ respectively where $i\in C_{s,k}$, \eqref{eq:P2} can be rewritten as the following optimization problem.
	With cluster assignment $C_{l,k}$ and sparse code $\Z_l$ fixed, we solve \eqref{eq:sub_pro_mOmega} to update $\omg_{l,k}$. 
	\begin{equation}\label{eq:sub_pro_mOmega}
		\begin{aligned}
			&\min_{\{\omg_{l,k}\}} \sum_{s=l}^L \sum_{k=1}^{K_s}  \bigg\{  \|\omg_{s,k}\mathbf{R}_{s,C_{s,k}} - \Z_{s,C_{s,k}}\|_\mathrm{F}^2\bigg\} \;\;\quad \mathrm{s.t.} \,\, \omg_{l,k} \omg_{l,k}^\top = \omg_{l,k}^\top \omg_{l,k}  = \I, \forall l,k.
		\end{aligned}
	\end{equation}
	
	Define $\B^{l\leftarrow s}_{C_{l,k}}$ as the matrix consisting of back-propagation information vectors from the $s$-th layer to the $l$-th layer for all patches in layer $l$ belonging to cluster $k$. We need to handle the same entanglement between transforms in the former layers and later layers. With the same argument to untwine such dependence as in equation \eqref{eq:disentangle}, the objective function in problem \eqref{eq:sub_pro_mOmega} can be re-written as
\begin{equation}\label{eq:sub_pro_mOmega_3}
	\begin{aligned}
		&\min_{\{\omg_{l,k}\}} \sum_{k=1}^{K_l}  \bigg\{  \|\omg_{l,k}\R_{l,C_{l,k}} - \Z_{l,C_{l,k}}\|_\mathrm{F}^2\bigg\} + \sum_{s=l+1}^L \sum_{k=1}^{K_l} \bigg\{  \|\omg_{l,k}\R_{l,C_{l,k}} - \Z_{l,C_{l,k}} - \B^{l\leftarrow s}_{C_{l,k}}\|_{\mathrm{F}}^2\bigg\} 
		\\
		&\sim \min_{\{\omg_{l,k}\}} \sum_{k=1}^{K_l}  \bigg\{\| \omg_{l,k}\R_{l,C_{l,k}} - \Z_{l,C_{l,k}} - \frac{1}{L-l+1} \sum_{s=l+1}^L \B^{l\leftarrow s}_{C_{l,k}}\|_{\mathrm{F}}^2 \bigg\} ,
	\end{aligned}
\end{equation}
where the symbol $\sim$ means equal up to some additive terms that do not depend on the variables being optimized for. 



From the rewritten objective function, the unitary constraint on transforms reduces the optimization over each $\omg_{l,k}$ to the orthogonal procrustes problem, which thus can be solved separately. Define $\mathbf{G}_{l,k}$ as
\begin{equation}\label{eq:mOmega_G_kl}
	\mathbf{G}_{l,k} =
	\begin{cases}
		\R_{l,C_{l,k}}\big( \Z_{l,C_{l,k}} + \frac{1}{L-l+1} \sum_{s=l+1}^L \B_{C_{l,k}}^{l\leftarrow s} \big)^{\top}, & 1\leq l \leq L-1,\\
		\R_{L,C_{L,k}}\Z_{L,C_{L,k}}^{\top}, & l=L.
	\end{cases}
\end{equation}

In the $l$-th layer, the solution to each orthogonal procustes problem (i.e.\ the solution to \eqref{eq:sub_pro_mOmega}) 
with respect to each transform $\omg_{l,k}$ is 
\begin{equation}\label{eq:mOmega_l_sol}
	\widehat{\omg}_{l,k}= \V_{l,k}\U_{l,k}^\top,
\end{equation}
where $\U_{l,k}$ and $\V_{l,k}$ are the singular vector matrices in the singular value decomposition of the matrix $\mathbf{G}_{l,k} = \U_{l,k}\Sig_{l,k}\V_{l,k}^\top$~\cite{orthoProcrustes_66}.
\end{itemize}

\begin{algorithm}[H]  
\caption{MARS Learning Algorithm~\cite{yang:20:mars}}\label{alg: mcst_train}
\begin{algorithmic}[0]
	\State \textbf{Input:}
	training data $\R_1$, all-zero initial $\{\widetilde{\Z}_l^{(0)}\}$, initial $\{\widetilde{\omg}_{1,k}^{(0)}, 1\leq k \leq K_1\} =$ 2D DCT, random matrices for initial $\{\widetilde{\omg}_{l,k}^{(0)}, 2\leq l \leq L, 1\leq k \leq K_l\}$, k-means initialization for the $1$st layer cluster assignment $\{\widetilde{C}_{1,k}^{(0)}, 1\leq k \leq K_1\}$ and random initialization for cluster assignment in the rest of layers, thresholds $\{\eta_l\}$, number of iterations $T$.			
	\State \textbf{Output:}  learned transforms $\{\widetilde{\omg}_{l,k}^{(T)}, 1\leq l \leq L, 1\leq k \leq K_l\}$.
	\For {$t =1,2,\cdots,{T}$}		
	\For {$l =1,2,\cdots,{L}$}	
	\State \textbf{1)} Cluster assignment update for $\{\widetilde{C}_{l,k}^{(t)},1\leq k \leq K_l\}$ by solving \eqref{eq:update_Cl}.	
	
	\State \textbf{2)} Sparse Coding for $\widetilde{\Z}_l^{(t)}$ via \eqref{eq:sol_Z_l}.	
	
	\State \textbf{3)} Updating $\{\widetilde{\omg}_{l,k}^{(t)}, 1\leq k \leq K_l\}$ via \eqref{eq:mOmega_l_sol}.
	
	\EndFor	
	\EndFor	
\end{algorithmic}
\end{algorithm}

\subsubsection{PWLS-MCST image reconstruction algorithm}

We adopt the Penalized Weighted Least Squares (PWLS) approach to reconstruct an image from its noisy sinogram. The learned MCST model is incorporated into the cost function as the regularizer term. The specific PWLS problem is shown in~\eqref{eq:P3}, where $\x \in \mathbb{R}^{N_p}$ denotes the reconstructed image and $N_p$ is the number of pixels in the image. Vector  $\y \in \mathbb{R}^{N_d}$ represents the noisy sinogram data and $\A\in \mathbb{R}^{N_d\times N_p}$ is the system matrix of the CT scan. $\W=\diag \{w_i\} \in \mathbb{R}^{N_d \times N_d}$ is the diagonal weighting matrix with $w_i$ indicating the inverse variance of $y_i$. Matrix $\H_{\A}$ is the diagonal majorizing matrix of $\A^T\W\A$. Operator $\P_{i}$ extracts the $i$-th overlapping patch from the original image with patch stride of $1$ pixel. Parameter $\beta$ controls the trade-off between the data-fidelity term and the regularizer term. $\{\gamma_l\}$ are tunable thresholds which control the sparsity of the MCST model in each layer.

\begin{equation}\label{eq:P3}	
\min_{\x \geq \0}  \frac{1}{2}\|\y - \A \x\|^2_{\W}  + \beta \Rsf(\x),   
\tag{P1}
\end{equation}
where $\Rsf(\x)$ is defined as
\begin{equation*}\label{eq:Rx_MCST}
\begin{aligned}
		&\Rsf(\x) \triangleq  \min_{\{\Z_l, C_{l,k}\}}   \sum_{l=1}^L \sum_{k=1}^{K_l} \sum_{i\in C_{l,k}} \bigg\{  \|\omg_{l,k}\r_{l,i}(\x) - \z_{l,i}\|_2^2 + \gamma_l^2\|\z_{l,i}\|_0  \bigg\} \\
		& \mathrm{s.t.} \quad \begin{cases}
			\r_{1,i}(\x)  = \P_{i}\x, & \forall \, i, \\
			\r_{l+1,i}(\x) = \omg_{l,k}\r_{l,i}(\x) - \z_{l,i}, & \forall i, \,\, 1\leq l\leq L-1.
		\end{cases}
	\end{aligned}
\end{equation*} 

To solve \eqref{eq:P3}, we decompose \eqref{eq:P3} into several subproblems (image update for $\x$, cluster assignment update for $\{C_{l,k},1\leq l\leq L, 1\leq k \leq K_l\}$, sparse coding for $\{\Z_l,1\leq l \leq L\}$) and solve these subproblems sequentially in each iteration of the proposed algorithm. The complete algorithm corresponding to the image reconstruction stage is shown in \textbf{Algorithm} \ref{alg: mcst_recon}. 

\begin{itemize}
	\item [a)] Image update step
	
	Here, we fix the cluster assignments $\{C_{l,k},1\leq l\leq L, 1\leq k \leq K_l\}$ and sparse representations $\{\Z_l, 1\leq l \leq L\}$. The resulting subproblem can be rewritten as follows.
	
	\begin{equation}\label{eq:img_update}	
		\min_{\x \geq \0}  \frac{1}{2}\|\y - \A \x\|^2_{\W}  + \beta \Rsf_2(\x)
	\end{equation}
	\begin{equation*}\label{eq:Rx_img_update}
		\begin{aligned}
			&\Rsf_2(\x) \triangleq  \sum_{l=1}^L \sum_{k=1}^{K_l} \sum_{i\in C_{l,k}} \bigg\{  \|\omg_{l,k}\r_{l,i}(\x) - \z_{l,i}\|_2^2 \bigg\}
			\\
			&\mathrm{s.t.} \quad \begin{cases}
				\r_{1,i} (\x)= \P_{i}\x, & \forall \, i \\
				\r_{l+1,i}(\x) = \omg_{l,k}\r_{l,i}(\x) - \z_{l,i}, & \forall i,\, 1\leq l\leq L-1,.
			\end{cases}
		\end{aligned}
	\end{equation*} 
	
	The solving process for~\eqref{eq:img_update} is identical to the image update step in Ref.~\cite{yang:20:mars}
	For each outer iteration, we update the image $T_i$ times with parameter $\rho$ decreasing as in \eqref{eq:rho_decrease}. The formulae to compute $\nabla\Rsf_2(\x)$ and the Hessian matrix $\H_{\Rsf_2}$ for image update are given in \eqref{eq:grad_R} and \eqref{eq:D_R}, respectively. 
	
	\begin{equation}\label{eq:rho_decrease}		
		\rho_r(\alpha) =  
		\begin{cases}
			1, & r=0,\\
			\frac{\pi}{\alpha(r+1)}\sqrt{1-(\frac{\pi}{2\alpha(r+1)})^2}, & \text{otherwise},
		\end{cases}\\
	\end{equation}
	
	\begin{equation}\label{eq:grad_R}		
		\nabla \Rsf_2(\x)= 2 \beta \sum_{i=1}^{N_p} (\P_{i})^\top \bigg\{ L\P_{i}\x - \sum_{k=1}^L \b^{0\leftarrow k}_{i} \bigg\},
	\end{equation}	
	\begin{equation}\label{eq:D_R}		
		\H_{\Rsf_2}  \triangleq  \nabla^2 \Rsf_2(\x) = 2L \beta \sum_{i=1}^{N_p}  (\P_{i})^\top\P_{i}, \\
	\end{equation}
	
	\item [b)] Cluster assignment update for $\{C_{l,k}, 1\leq l \leq L, 1 \leq k \leq K_l\}$
	
	In the process of image reconstruction, the cluster assignment for every reconstructed image patch is updated dynamically in the same manner as in the learning stage. We find the optimal class for the $i$-th patch in the $l$-th layer 
	by solving the following problem as in Section~\ref{sec:alg_mcst_train}.
	\begin{equation}\label{eq:update_Cl_recon}
		\begin{aligned}
			&\widehat{k}(i,l) = \arg\min_{1\leq k \leq K_l}\|\omg_{l,k}\r_{l,i} - \z_{l,i}\|_2^2 + \sum_{s=l+1}^L \|\omg_{s,k(i,s)}\r_{s,i} - \z_{s,i}\|_2^2 + \eta_l^2\|\z_{l,i}\|_0  
			\\
			& \quad \mathrm{s.t.} \quad 
			\begin{cases}
				\r_{1,i} &= \P_{i}\x, \\
				\r_{s,i} &= \omg_{s-1,k(i,s-1)}\r_{s-1,i} - \z_{s-1,i}, \,\,  l+1\leq s\leq L-1.
			\end{cases}
		\end{aligned}
	\end{equation}
	
	\item [c)] Sparse coding step for $\{\Z_l,1\leq l \leq L\}$
	
	In this step, by eliminating variables except the target sparsecode $\Z_l$, \eqref{eq:P3} gets simplified to~\eqref{eq:sub_pro_Z_recon} which is the same as the sparse coding problem in \eqref{eq:mcst_sub_pro_Z}. 
	\begin{equation}\label{eq:sub_pro_Z_recon}
		\begin{aligned}
			&\min_{\Z_l} \sum_{s=l}^L \sum_{k=1}^{K_s} \sum_{i\in C_{s,k}} \bigg\{  \|\omg_{s,k}\r_{s,i} - \z_{s,i}\|_2^2\bigg\} + \gamma_l^2\|\Z_{l}\|_0  
		\end{aligned}
	\end{equation}	
	The solution to \eqref{eq:sub_pro_Z_recon} is 
	given in \eqref{eq:sol_Z_l}.
	
\end{itemize}

\begin{algorithm}[!h]  
	\caption{Image Reconstruction Algorithm}\label{alg: mcst_recon}
	\begin{algorithmic}[0]
		\State \textbf{Input:}
		initial image $\widetilde{\x}^{(0)}$, all-zero initial $\{\widetilde{\Z}_l^{(0)}\}$, pre-learned $\{\omg_l\}$, thresholds $\{\gamma_l\}$, \\
		$\alpha = 1.999$, $\H_{\A}$, $\H_{\Rsf_2}$, number of outer iterations $T_{O}$,  number of inner iterations $T_i$.				
		\State \textbf{Output:}  reconstructed image $\widetilde{\x}^{(T_{O})}$.
		\For {$t =0,1,2,\cdots,{T_{O}-1}$}		
		
		\State \hspace{-0.10in}\textbf{1) Image Update}: With $\{\widetilde{\Z}_l^{(t)}\}$ fixed,	
		
		\hspace{-0.10in}\textbf{Initialization:} $\rho=1$, $\x^{(0)} = \widetilde{\x}^{(t)}$, $\g^{(0)} = \ze^{(0)}  = \A^{\top}\W(\A\x^{(0)}-\y) $ and $\h^{(0)} = \H_\A \x^{(0)} - \ze^{(0)}$.	
		\For {$r =0,1,2,\cdots,T_i-1,$}				
		\begin{equation*}
			\left\{			
			\begin{aligned}
				\s^{(r+1)} &= \rho(\H_\A \x^{(r)} -\h^{(r)}) + (1-\rho)\g^{(r)} \\
				\x^{(r+1)} &= [\x^{(r)} - (\rho\H_\A+\H_{\Rsf_2})^{-1}(\s^{(r+1)} +\nabla \Rsf_2(\x^{(r)}))]_+ \\
				\ze^{(r+1)} & \triangleq  \A^\top\W(\A\x^{(r+1)}-\y)   \\
				\g^{(r+1)} &= \frac{\rho}{\rho+1}(\alpha \ze^{(r+1)} + (1-\alpha)\g^{(r)}) +  \frac{1}{\rho+1}\g^{(r)}\\
				\h^{(r+1)}  &= \alpha(\H_{\A} \x^{(r+1)} -\ze^{(r+1)}) + (1-\alpha)\h^{(r)} 
			\end{aligned}
			\right.
		\end{equation*}  
		\State decreasing $\rho$ using \eqref{eq:rho_decrease}. 
		\EndFor	
		\State   $\widetilde{\x}^{(t+1)} = \x^{(T_i)}$. 
		\State \hspace{-0.10in}\textbf{2) Cluster assignment update}: with $\widetilde{\x}^{(t+1)}$ and sparse code $\{\Z_l, 1\leq l \leq L\}$ fixed, for each $1\leq l \leq L$, update $\{C_{l,k},1\leq k \leq K_l \}$ sequentially by \eqref{eq:update_Cl_recon}.
		\State \hspace{-0.10in}\textbf{3) Sparse Coding}: with $\widetilde{\x}^{(t+1)}$ and cluster assignment $\{C_{l,k},1\leq l \leq L, 1\leq k \leq K_l\}$ fixed, for each $1\leq l \leq L$, update $\widetilde{\Z}_l^{(t+1)}$ sequentially by \eqref{eq:sol_Z_l}.
		\EndFor	
	\end{algorithmic}
\end{algorithm}


\section{Experiments}\label{chap:MCST_experiment}

\subsection{Experiment setup}
In this study, we evaluate the performance of the proposed MCST sparse signal model and the PWLS-MCST algorithm for LDCT reconstruction. We investigate the PWLS-MCST algorithm with different numbers of layers (2, 3 layers) with each layer containing 5 clusters to classify input patches into various groups. We compare the proposed methods against several baseline methods including FBP~\cite{feldkamp:84:pcb}, PWLS with EP regularization~\cite{cho:15:rdf}, and the recent MARS~\cite{yang:20:mars} and ULTRA~\cite{zheng:18:pua} regularizations in PWLS, respectively. The details of these methods are described as follows.
\begin{enumerate}
	\item \textbf{FBP}: We used a Hanning window with $0.4$ width to accomplish the back-projection reconstruction.
	\item \textbf{PWLS-EP}: An edge-preserving (EP) regularizer is employed in the  PWLS reconstruction scheme. The mathematical representation of the EP regularizer is $\mathsf{R}(\x)=\sum_{j=1}^{N_p} \sum_{k\in N_{j}}\kappa_{j} \kappa_{k} \phi(x_j - x_k)$, where $N_p$ counts the overall number of pixels, $N_j$ denotes the neighborhood of the $j$th pixel
	, and $\kappa_{j}$ and $\kappa_{k}$ denote the analytically determined parameters to encourage uniform resolution. 
	$\phi(t) \triangleq \delta^2(\sqrt{1+|t/\delta|^2}-1)$ 
	is a potential function.
	\item \textbf{PWLS-MARS}: A multi-layer residual sparsifying transform (no clustering involved) is adopted as the prior in PWLS reconstruction. For a fair comparison, we choose MARS models with $2$, $3$ layers (MARS2, MARS3), which have the same depth as the MCST models in our experiments. 
	\item \textbf{PWLS-ULTRA}: PWLS with a union of transforms model, which is equivalent to the MCST signal model with a single layer.
\end{enumerate}

We used two metrics, RMSE and SSIM~\cite{xu:12:ldx,zhang:17:tbd}, to evalute the quality of reconstructions. We compute the root mean square error (RMSE) as RMSE $= \sqrt{\Sigma_{i \in \text{ROI}}(\widehat{\x}_i-\x^\star_i)^2/{N_{\text{ROI}}}}$, where $\widehat{\x}$ and $\x^\star$ denote the reconstructed image and ground truth image, respectively, and $N_{ROI}$ calculates the pixel number inside the region of interest (ROI), which is a circular region containing all structures and tissues.

Our experiments include two main parts: learning transforms for constructing regularizers, and using them for low-dose CT image reconstruction. We work with two datasets: XCAT phantom simulated data~\cite{Segars:08:rcs} and Mayo clinic data~\cite{Mayo:16:data}. We generate the low-dose measurements with the ``Poisson + Gaussian'' noise model, i.e., $\widehat{\bf{y}}_i = \mathrm{Poisson} \{I_0 \mathrm{e}^{-[\bf{Ax}]_i}\} + \mathcal{N}\{0, \sigma^2\}$~\cite{ding:16:mmp}, where $I_0$ denotes the incident photon intensity of the X-ray beam and we set $I_0=1\times 10^4$ per ray and with no scatter. $\sigma^2=5^2$ is the variance of electronic noise. 
For XCAT phantom data, we use 5 slices to train the model. We simulate the low-dose measurements with GE 2D LightSpeed fan-beam geometry. The size of measurements is $888\times 984$. The pixel size for the XCAT phantom is $\Delta_x=\Delta_y=0.4883$mm. 
For the Mayo Clinic data, 7 regular-dose images collected from three patients were used to train the MCST model. We synthesized the low-dose measurements using a fan-beam CT geometry with a monoenergetic source. The sinograms are of size $736\times 1152$. Specifically, the width of the detector column is $1.2858$mm. The distances from the source to detector and to rotation center are $1085.6$mm and $595$mm, respectively. The size of the reconstructed images is  $512\times 512$ with $\Delta_x = \Delta_y = 0.69$mm.

\subsection{Low-dose experiments with XCAT phantom dataset}
\subsubsection{Behavior of the learned MCST models}
\begin{figure}[!h]
	\centering
	\includegraphics[width=0.8\textwidth]{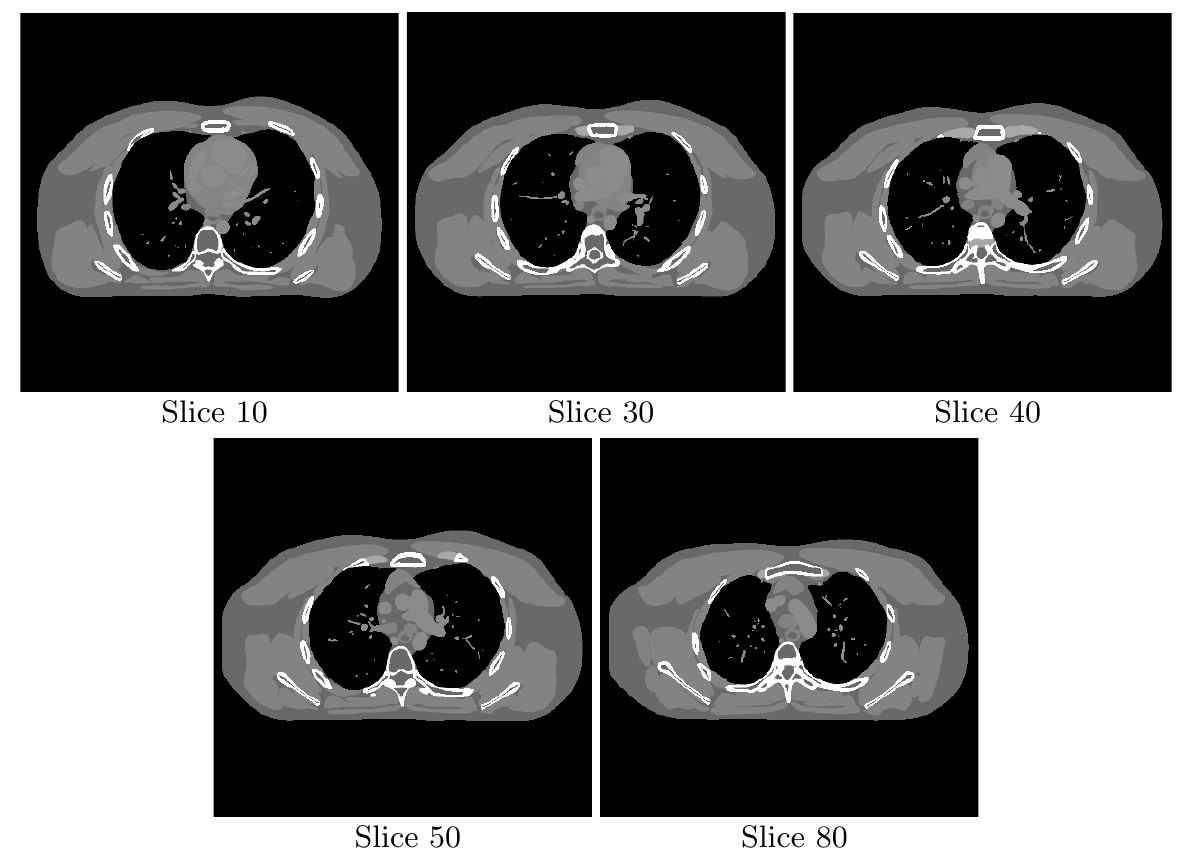}
	\caption{XCAT phantom slices for training MCST models. The number below each subfigure indicates its location in the volume.}
	\label{fig:XCAT_fig_train}
\end{figure}

First, we display the $64\times 64$ learned transforms from the XCAT phantom training dataset. 
Figure~\ref{fig:XCAT_fig_train} 
shows the image set with all training slices. The number below each subfigure indicates its location in the volume. We extracted overlapping $8\times8$ patches with a patch stride $1\times1$. The total number of training patches is approximately $8.5\times 10^5$. 

\begin{figure}[!h]
	\centering
	\includegraphics[width=1.0\textwidth]{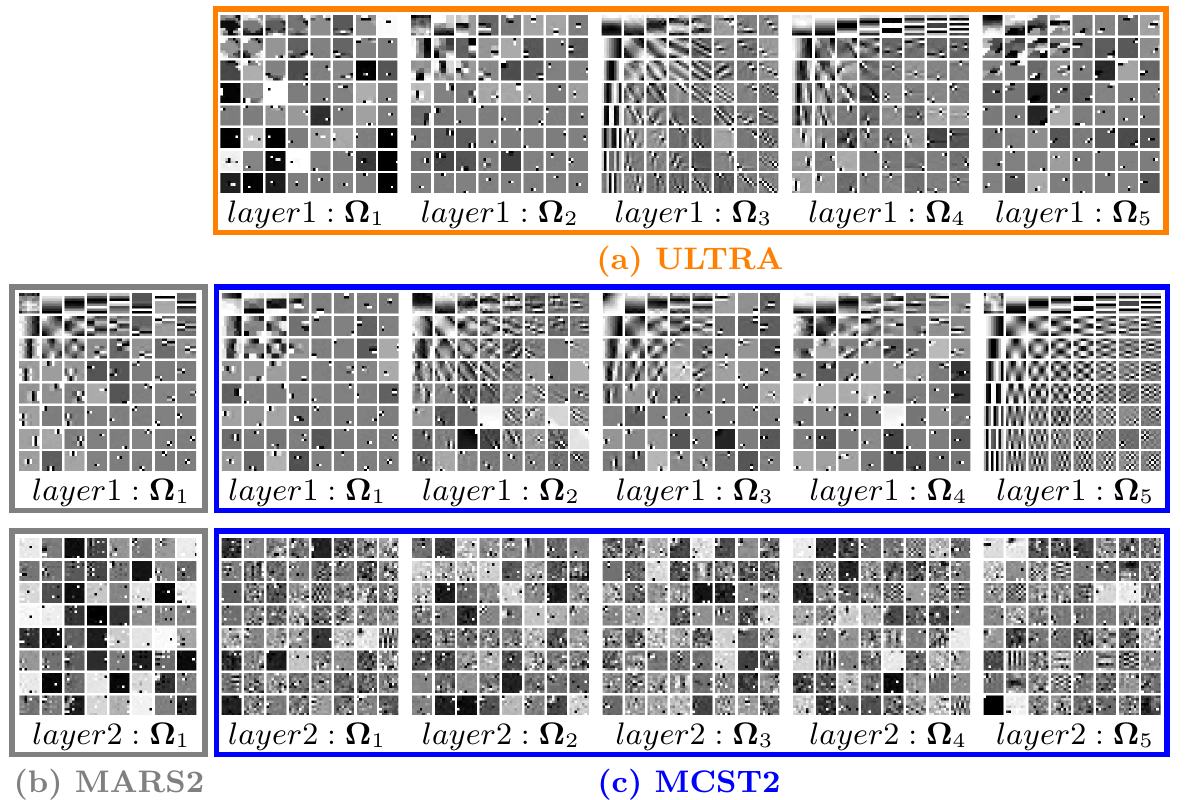}
	\caption{The pre-learned $64\times 64$ transforms of ULTRA with 5 clusters (shown in the orange box), MARS with two layers (shown in the gray box), and MCST2 with 5 clusters in both layers (shown in the blue box). Each row of the transforms is reorganized as an $8 \times 8$ square matrix for simplicity.}
	\label{fig:XCAT_trans}
\end{figure}

We learn the transforms in the MCST method with different model depth: ULTRA (single-layer MCST), MCST2 (2-layer), and MCST3 (3-layer), where each layer contains 5 clusters to enable learning 
rich features. To verify the effectiveness of the multi-layer and multi-cluster extension, MARS2 (2-layer) and MARS3 (3-layer) are included for comparison. 
We ran 1000 iterations to train each model and vary the parameters as for the MARS model in our previous work~\cite{yang:20:mars}. Specifically, we set parameters $\eta=80$ for ULTRA, $(\eta_1,\eta_2)=(80,60)$ for MCST2 and MARS2, $(\eta_1,\eta_2,\eta_3)=(90,80,60)$ for MARS3, $(\eta_1,\eta_2,\eta_3)=(80,60,40)$ for MCST3. 
We initialize the transforms for MCST in the first layer with 2D DCT matrices and the rest of the transforms are initialized with random matrices.
Figure~\ref{fig:XCAT_trans} shows the pre-learned $64\times 64$ transforms for ULTRA with 5 clusters (shown in the orange box), MARS with two layers (shown in the gray box), and MCST2 with 5 clusters in both layers (shown in the blue box). Each row of the learned transforms is reorganized as an $8 \times 8$ square matrix for display purposes. The results show that the MCST model integrates the advantages of ULTRA model and MARS model. In particular, the transforms in the first layer of MCST show rich features (like ULTRA), while the transforms in the second layer capture finer features by further sparsifying the representation of residuals. 

\subsubsection{LDCT reconstruction results with PWLS-MCST algorithm}
\begin{figure}[!h]
	\centering
	\includegraphics[width=1.0\textwidth]{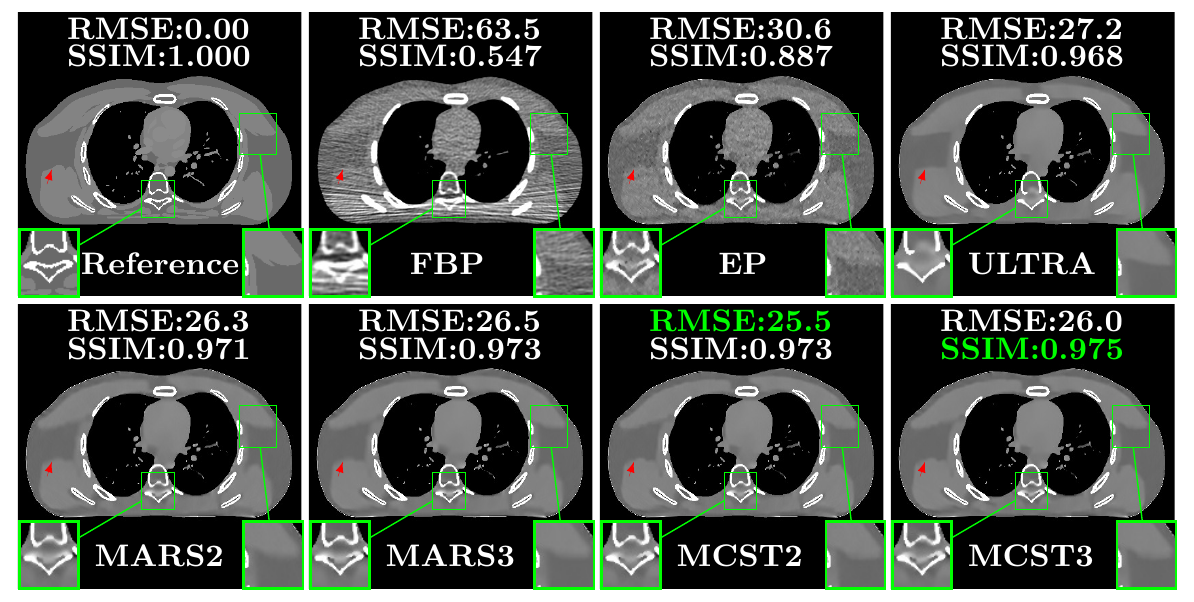}
	\caption{Comparison of reconstructions of slice 20 of the XCAT phantom with FBP, PWLS-EP, PWLS-ULTRA, PWLS-MARS2, PWLS-MARS3, PWLS-MCST2, and PWLS-MCST3, respectively. The display window is $[800,1200]$ HU.}
	\label{fig:XCAT_slice20_recon}
\end{figure}

\begin{figure}[!h]
	\centering
	\includegraphics[width=1.0\textwidth]{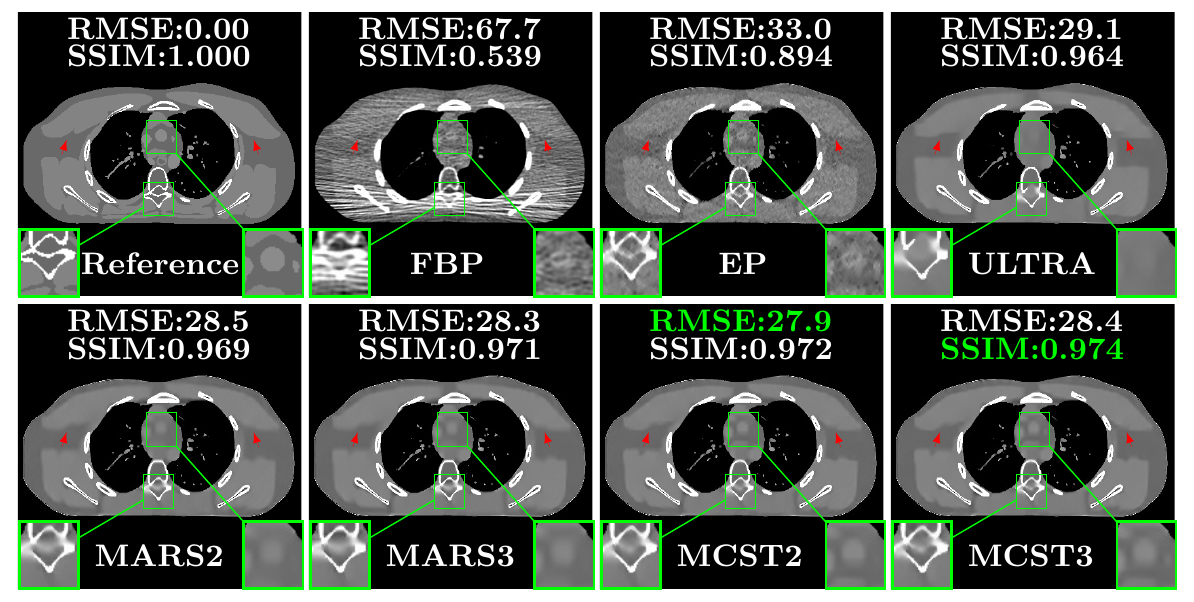}
	\caption{Comparison of reconstructed slice 60 of the XCAT phantom with FBP, PWLS-EP, PWLS-ULTRA, PWLS-MARS2, PWLS-MARS3, PWLS-MCST2, and PWLS-MCST3, respectively. The display window is $[800,1200]$ HU.}
	\label{fig:XCAT_slice60_recon}
\end{figure}

The reconstructed slice $20$ and slice $60$ of the XCAT phantom dataset are illustrated in Figures~\ref{fig:XCAT_slice20_recon} and \ref{fig:XCAT_slice60_recon}, respectively. We compare the results of PWLS-MCST with the conventional method FBP, PWLS-EP, PWLS-ULTRA, and PWLS-MARS. We used the FBP reconstruction as initialization for PWLS-EP and set the regularization parameter as $\beta=2^{16}$. We ran 1000 iterations of the relaxed LALM algorithm for PWLS-EP and ran 1500 outer iterations for other ST-based iterative methods to ensure their convergence. We take the same training dataset and validate image slice (slice 48 of the XCAT phantom) as in Ref.~\cite{yang:20:mars}. The reconstruction parameters $\{\beta,\gamma_l\}$ are varied based on their values in Ref.~\cite{yang:20:mars}. Specifically, we set the regularization parameter $\beta$ and sparsity parameters $\{\gamma_l\}$ as $(\beta,\gamma) = (2\times 10^5, 30)$ for ULTRA, $(\beta,\gamma_1,\gamma_2) = (9\times 10^4,30,10)$ for MARS2 and MCST2, $(\beta,\gamma_1,\gamma_2,\gamma_3) = (9\times 10^4, 25,15,10)$ for MARS3, and $(\beta,\gamma_1,\gamma_2,\gamma_3) = (9\times 10^4, 30,12,10)$ for MCST3, respectively. 

The reconstruction results show that PWLS-MCST attains better performance than all baseline methods. Compared with PWLS-MARS, the images reconstructed by PWLS-MCST demonstrate greater contrast 
due to flexible image modeling with
multi-class transforms learned from different groups of patches. The performance difference is exemplified in the zoomed-in portions and details pointed by red arrows. Also, using multiple layers enables PWLS-MCST to preserve more key features compared to PWLS-ULTRA. For example, in the reconstructed images by PWLS-ULTRA, some subtle bone structures are missing (zoomed-in region at the bottom-left corner of subfigures), whereas PWLS-MCST mitigates this effect; the zoom-in regions located at the bottom-right corner of subfigures for PWLS-MCST are closer to the ground truth than those with PWLS-ULTRA. We point out that PWLS-MCST provides the best performance in terms of RMSE and SSIM.

However, compared with the 2-layer MCST model, the improvement with deeper MCST model (MCST3) is limited, which is in accordance with the performance of PWLS-MARS. Since XCAT phantom images have relatively simple structures, the complexity of MCST3 models may be too high to demonstrate significant usefulness with this dataset. A larger number of tunable parameters in deeper MCST models might also create potential sub-optimality in the training process.

\subsection{Low-dose experiments with Mayo Clinic dataset}

\subsubsection{MCST model training}

\begin{figure}[!h]
	\centering
	\includegraphics[width=1.0\textwidth]{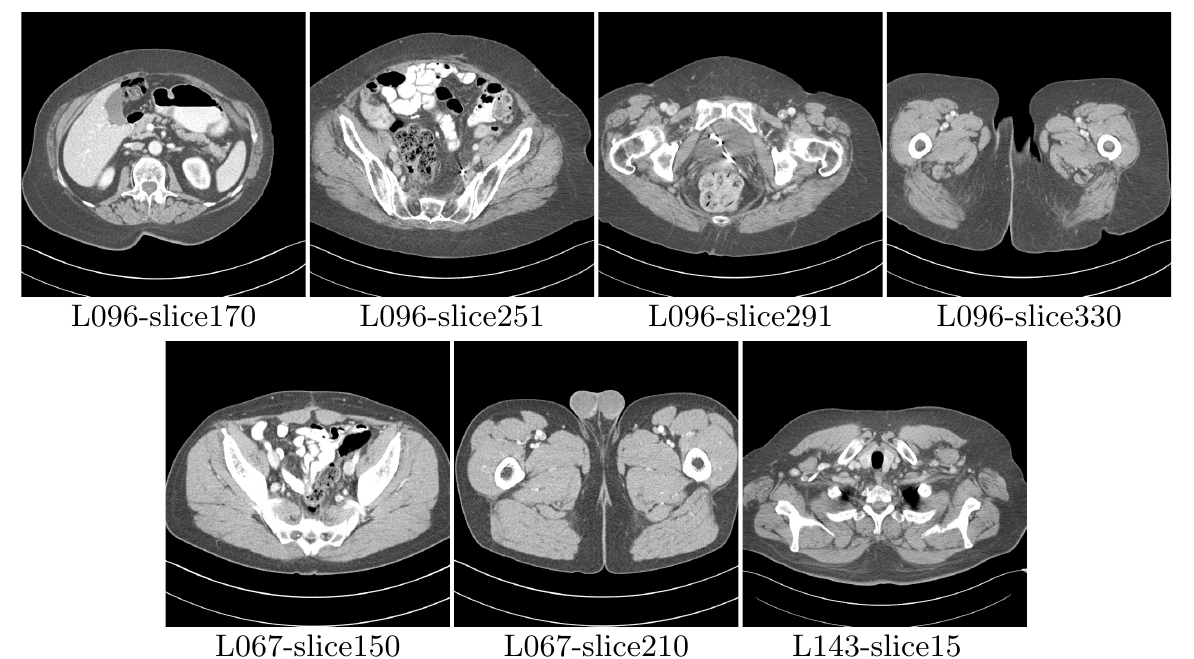}
	\caption{Training dataset selected from Mayo Clinic data. Seven regular-dose slices collected from patients L096, L067, L143 are used to train the MCST model.}
	\label{fig:Mayo_fig_train}
\end{figure}

Next, we train the MCST model on
the
Mayo Clinic dataset. We use 7 regular-dose slices collected from three different patients (L096, L067, L143) for training. 
Figure~\ref{fig:Mayo_fig_train} 
displays all training slices used in the experiment. Similar to the XCAT phantom experiment, we extracted $8\times 8$ overlapping patches from these training slices. The total number of patches is $\approx 1.8 \times 10^6$. 

We trained the ST-based signal model with similar parameters as for the XCAT phantom experiment. We set iteration number for training as 1000 and set $\eta=80$ for ULTRA, $(\eta_1,\eta_2) = (80,60)$ for MARS2 and MCST2, $(\eta_1,\eta_2,\eta_3) = (60,60,40)$ for MARS3, and $(\eta_1,\eta_2,\eta_3) = (80,60,40)$ for MCST3. The initialization of the transforms in the MCST model are DCT matrices for the first layer  and random matrices for successive layers, respectively.

\begin{figure}[!h]
	\centering
	\includegraphics[width=1.0\textwidth]{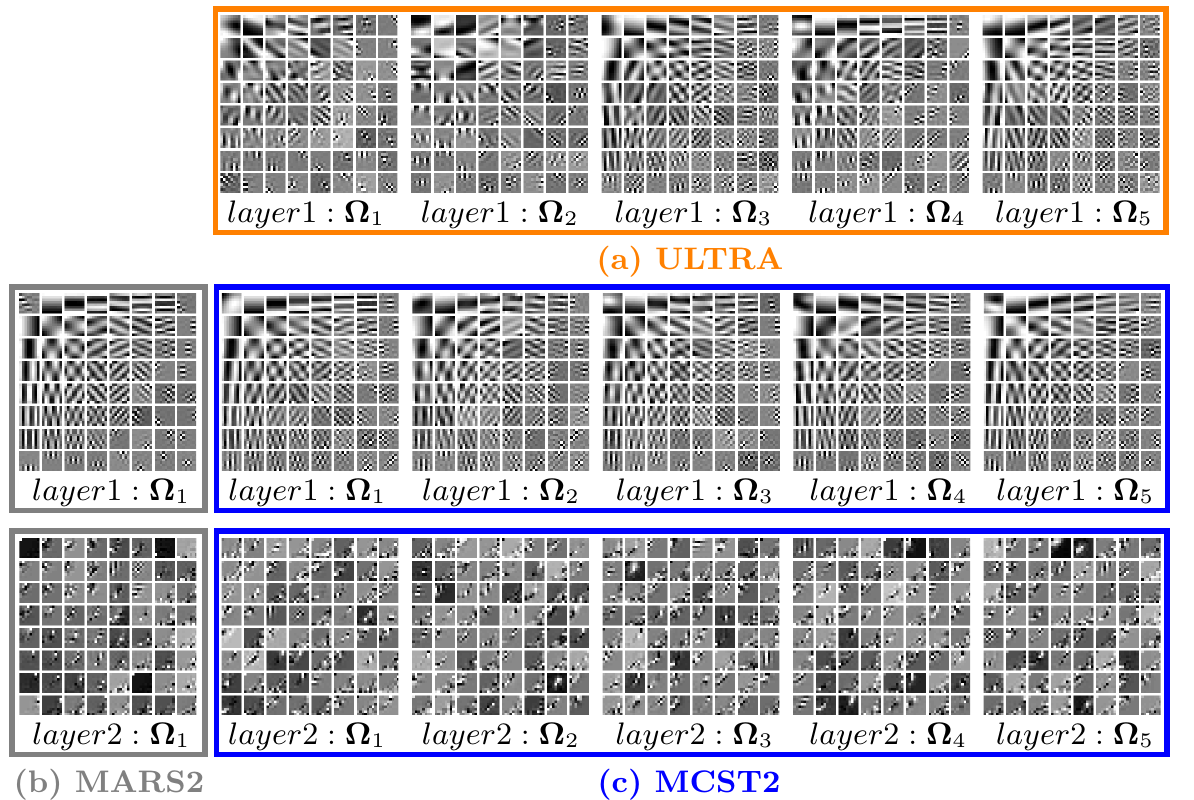}
	\caption{Learned transforms for the Mayo Clinic data. The transforms in the orange box are obtained from the ULTRA learning algorithm, while the transforms in the gray and blue boxes correspond to the MARS2 and MCST2 model, respectively.}
	\label{fig:Mayo_trans}
\end{figure}

As Figure~\ref{fig:Mayo_trans} shows, the learned transforms from the ULTRA learning algorithm contain some compound features (e.g. the third transform in the first row). Similar to the XCAT phantom experiment, MCST2 blends the advantages of MARS with the ULTRA model and captures richer features from the training patch set. Furthermore, considering the complexity of the Mayo Clinic dataset, MCST2 is able to learn effective features in every transform of the first layer. 

\subsubsection{Numerical and visual results on Mayo Clinic data}

After obtaining a series of transforms, we incorporate those pre-learned transforms into the PWLS scheme and compare various reconstruction methods. For PWLS-EP, we iterate the relaxed LALM algorithm with 1000 iterations with the FBP reconstruction as initialization. The hyperparameter $\beta$ is set as $2^{15.5}$. For PWLS-ULTRA, PWLS-MARS2, and PWLS-MARS3, we set the maximum outer iterations $T_O$ as 1500 and inner image update times $T_i=2$. Similar to the MARS work~\cite{yang:20:mars}, we tuned the parameters on one validation image (slice 100 of patient L506) and tested the performance of our proposed algorithm on other slices (slice 140 of patient L333, slice 90 of patient L109, and slice 120 of patient L067). We varied the reconstruction parameters based on the fine-tuned parameters in the PWLS-MARS work. Specifically, we chose $(\beta,\gamma)=(2.5\times 10^4,30)$ for PWLS-ULTRA, $(\beta,\gamma_1,\gamma_2)=(1.8\times 10^4,30,10)$ for PWLS-MARS2 and PWLS-MCST2, and $(\beta,\gamma_1,\gamma_2,\gamma_3)=(1.8\times 10^4,30,12,10)$ for PWLS-MARS3 and PWLS-MCST3, respectively.

\begin{figure}[h!]
	\centering
	\includegraphics[width=1.0\textwidth]{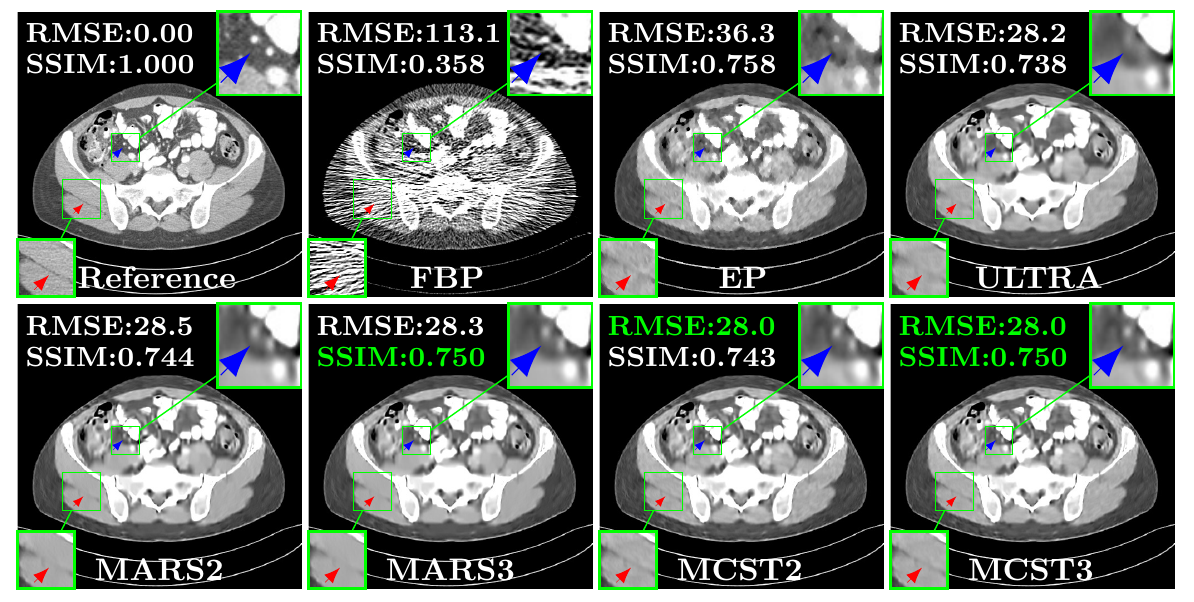}
	\caption{Comparison of the reconstructed slice 140 of patient L333 in the Mayo Clinic data with FBP, PWLS-EP, PWLS-ULTRA, PWLS-MARS2, PWLS-MARS3, PWLS-MCST2, and PWLS-MCST3 schemes, respectively. The display window is $[800,1200]$ HU.}
	\label{fig:Mayo_L333_slice140_recon}
\end{figure}

\begin{figure}[h!]
	\centering
	\includegraphics[width=1.0\textwidth]{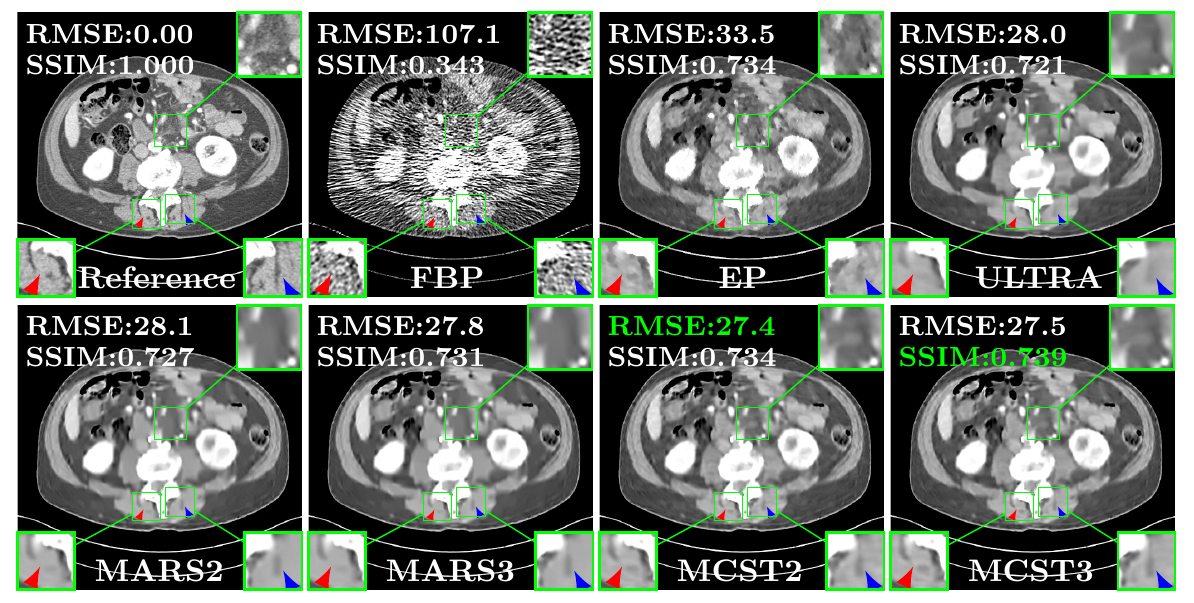}
	\caption{Comparison of the reconstructed slice 90 of patient L109 in the Mayo Clinic data with FBP, PWLS-EP, PWLS-ULTRA, PWLS-MARS2, PWLS-MARS3, PWLS-MCST2, and PWLS-MCST3 schemes, respectively. The display window is $[800,1200]$ HU.}
	\label{fig:Mayo_L109_slice90_recon}
\end{figure}

\begin{figure}[h!]
	\centering
	\includegraphics[width=1.0\textwidth]{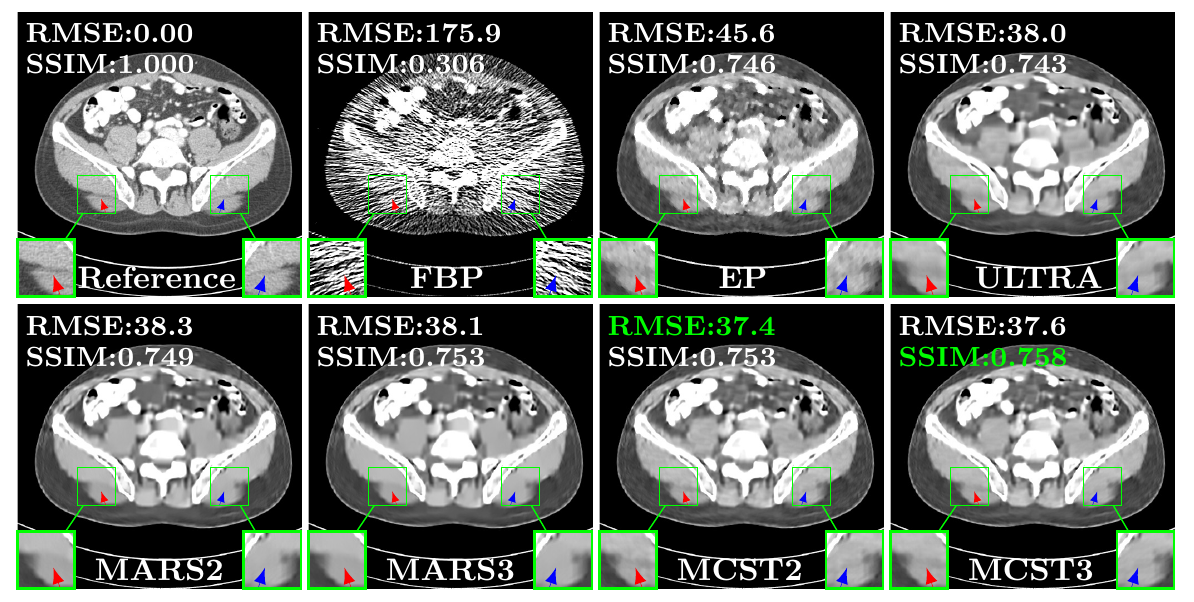}
	\caption{Comparison of the reconstructed slice 120 of patient L067 in the Mayo Clinic data with FBP, PWLS-EP, PWLS-ULTRA, PWLS-MARS2, PWLS-MARS3, PWLS-MCST2, and PWLS-MCST3 schemes, respectively. The display window is $[800,1200]$ HU.}
	\label{fig:Mayo_L067_slice120_recon}
\end{figure}

Figures~\ref{fig:Mayo_L333_slice140_recon}, \ref{fig:Mayo_L109_slice90_recon}, and~\ref{fig:Mayo_L067_slice120_recon} give the image reconstruction results of slice 140 of patient L333, slice 90 of patient L109, and slice 120 of patient L067, respectively. The proposed PWLS-MCST algorithm gives better quality of image reconstructions compared to other recent unsupervised learning-based methods (PWLS-ULTRA and PWLS-MARS), especially for recovering subtle details. Although the result of PWLS-ULTRA 
is clearer than the initial PWLS-EP reconstruction, it can be improved by incorporating 
the rich multi-layer structure from MCST. Numerical results indicate that PWLS-MCST performs the best in terms of RMSE and SSIM criteria.

\section{Conclusion}\label{chap:MCST_conclude}

In this work, we proposed \textit{a multi-layer clustering-based residual sparsifying transform} (MCST) network comprised of multiple sparsifying transform modules to extract rich information from training images. The MCST model is different from the previous single sparsifying transform model in two aspects. First, learning multiple transforms (in each layer) by clustering the data allows capturing rich features from patches. Second, the multi-layer structure aims to better capture the sparsity of the signal representation across several layers. 
We applied the learned MCST model to LDCT reconstruction and presented a novel image reconstruction algorithm dubbed PWLS-MCST. Experimental results indicate that PWLS-MCST provides better image reconstruction quality and image features than conventional methods such as FBP and PWLS-EP. Furthermore, PWLS-MCST shows advantages over several recent sophisticated algorithms for LDCT reconstruction including PWLS-MARS and PWLS-ULTRA, especially for displaying clearer edges and preserving subtle details. In the future, we plan to 
involve more sophisticated architectures like skipped connections among different layers and multi-scale transformation through down-sampling and up-sampling within the MCST model and formulations. Additional imaging applications can be explored to validate the general applicability of the proposed model.


\section*{Appendix: Notations}
We use lower case bold font to denote vectors and upper case bold font to denote matrices. $\|\cdot\|_2$ is the 2-norm for vectors, $\|\cdot\|_\mathrm{F}$ is the Frobenius norm for matrices, and $\|\cdot\|_0$ counts the non-zero entries. Operator $H_\eta(\cdot)$ is the hard-thresholding function which zeros out values less than $\eta$. We list the notation used in the description of our methodology in Table~\ref{table:Notation_sys_MCST}.

\begin{table}[h!]
	\centering
	\caption{Notation for MCST problem formulation}
	\begin{tabular}{c|c}
		\toprule
		$K_l$  & Number of classes in layer $l$\\
		\midrule
		$C_{l,k}$  &A set containing indices of patches in class $k$ in the $l$-th layer\\
		\midrule
		$\mathbf{r}_{l,i},\z_{l,i}$  & $i$-th column of the matrix $\R_l$ and $\Z_l$ \\
		\midrule
		$\R_{l,C},\Z_{l,C}$  & Submatrix of $\R_l$ and $\Z_l$ respectively, with column indices in the set $C$\\
		\midrule
		$\omg_{l,k}$ & Transform of the class $k$ in $l$-th layer\\
		\midrule
		$k(i,s)$ & Cluster assignment of the patch $i$ in $s$-th layer \\
		\midrule
		$\b^{p\leftarrow q}_{i}$  &  information back-propagation vector for the patch $i$ from $q$-th layer to $p$-th layer \\
		\midrule
		\makecell[c]{$\B_{C}^{p\leftarrow q}$} &  \makecell[c]{Information back-propagation matrix from $q$-th layer to $p$-th layer\\ for patches with indices in the set $C$}   \\
		\bottomrule
	\end{tabular}
	\label{table:Notation_sys_MCST}
\end{table}

\clearpage
\addcontentsline{toc}{section}{\numberline{}References}

\bibliographystyle{alpha}
\bibliography{refs.bib}

\end{document}